\input{aipcheck}

\documentclass[
    ,final            % use final for the camera ready runs
  ]
  {aipproc}

\layoutstyle{6x9}

\begin{document}

\title{Evolution of the symbiotic nova RX Puppis}

\author{J. Miko{\l}ajewska}{
  address={N. Copernicus Astronomical Center, Bartycka 18, 00716 Warsaw, Poland\\ 
mikolaj@camk.edu.pl}
}

\author{E. Brandi}{ address={Facultad de Ciencias Astron{\'o}micas y Geof\'{\i}sicas, UNLP -
 CIC - CONICET,  1900 La Plata, Argentina} }

\author{L. Garcia}{address={Facultad de Ciencias Astron{\'o}micas y Geof\'{\i}sicas, 
UNLP - CIC - CONICET,  1900 La Plata, Argentina}}

\author{O. Ferrer}{address={Facultad de Ciencias Astron{\'o}micas y Geof\'{\i}sicas, 
UNLP - CIC - CONICET,  1900 La Plata, Argentina}}

\author{C. Quiroga}{address={Facultad de Ciencias Astron{\'o}micas y Geof\'{\i}sicas, UNLP -
 CIC - CONICET,  1900 La Plata, Argentina}}

\author{G.C. Anupama}{address={Indian Institute of Astrophysics, Bangalore 560034, 
India}}

\begin{abstract} We present and discuss a hundred year history of activity of the hot 
component of RX Pup based on optical photometry and spectroscopy. The outburst evolution 
of  RX Pup resembles that of other symbiotic novae whereas at quiescence the hot component 
shows activity (high and low activity states) resembling   that of symbiotic recurrent 
novae T CrB and RS Oph. \end{abstract}

\maketitle

%%%%%%%%%%%%%%%%%%%%%%%%%%%%%%%%%%%%%%%%%%%%
%% MAINMATTER
%%%%%%%%%%%%%%%%%%%%%%%%%%%%%%%%%%%%%%%%%%%%

\section{Introduction}

RX Pup is a long-period interacting binary system   consisting of a Mira 
variable pulsating with P=578 days, surrounded by a thick dust shell, and a hot white 
dwarf companion accreting material from the Mira's wind. The binary separation could be as 
large as $a \geq 50$ a.u. (corresponding to $P_{\rm orb} \geq 200$ yr) as suggested by the 
permanent presence of a dust shell around the Mira component (\cite{m99}). 

\begin{figure} \includegraphics[width=\textwidth]{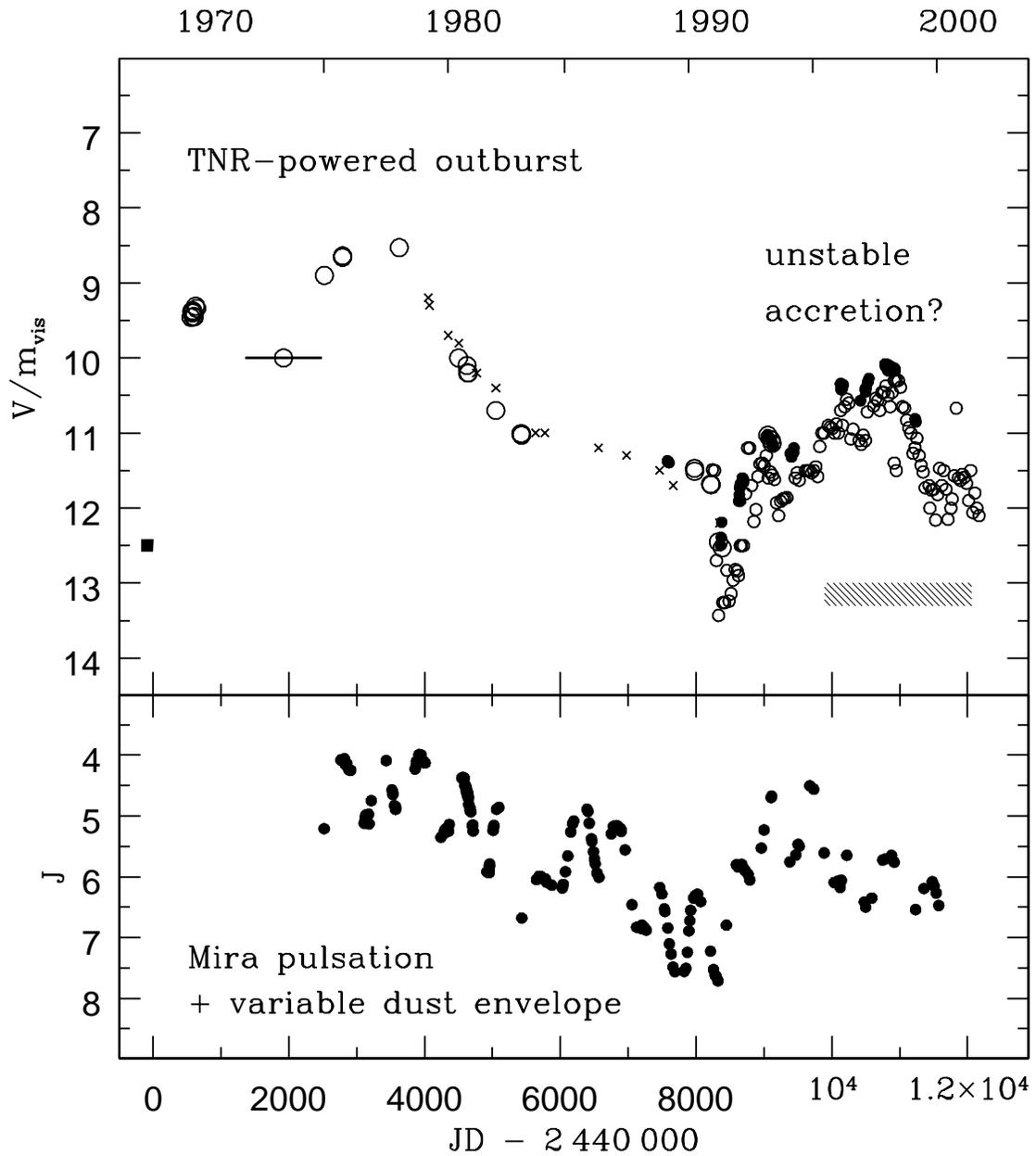} 
\caption{Optical and near- IR (J) light curves of RX Pup (\cite{m99}). In the $V/m_{\rm 
vis}$ light curve, small open circles represent RASNZ observations; large open circles and 
dots published V data; crosses FES magnitudes. The optical light curve is dominated by the 
hot component activity whereas the J light curve is dominated by the Mira pulsations and 
variable obscuration of the Mira by circumstellar dust. The shaded area indicates the 
period of our spectroscopic observations  discussed in Sec. 2.} \end{figure}

The analysis of multifrequency observations by \cite{m99} has shown that most, if not all, 
photometric and spectroscopic activity of RX~Pup in the UV, optical and radio range is due 
to activity of the hot component, while the Mira variable and its circumstellar 
environment is responsible for practically all changes in the IR range (Fig.1). In 
particular, RX Pup underwent a nova-like eruption during the last three decades. The 
evolution of the hot component in the HR diagram (Fig.11 of \cite{m99}) as well as 
evolution 
of 
the nebular emission in 1970$-$1993 is consistent with a symbiotic nova eruption, with the 
luminosity plateau reached in 1972/75 and a turnover in 1988/89. The hot component 
contracted in radius at roughly constant luminosity from c. 1972 to 1986; during this 
phase it was the source of a strong stellar wind and therefore could not accrete any 
further material. By 1991 the luminosity of the nova remnant had decreased to a few per 
cent of the maximum (plateau) luminosity, and the hot wind had practically ceased. By 1995 
the hot component start to accrete material from the Mira wind, as indicated by a general 
increase of the optical continuum and Balmer H\,{\sc i} emission. The quiescent optical 
spectrum of RX~Pup resembles the quiescent spectra of symbiotic recurrent novae, while the 
hot component luminosity is consistent with variable wind-accretion at a high rate, 
$\dot{M}_{\rm acc} \sim 10^{-7}\, \rm M_{\odot}\,yr^{-1}$ ($\approx 1$ per cent of 
$\dot{M}_{\rm cool}$). RX~Pup may be a recurrent nova; there is some evidence that a 
previous 
eruption occurred around 1894. 

In the following we discuss results of optical and red spectroscopic observations of RX 
Pup obtained during 1995$-$2001 with the REOSC echelle spectrograph at the 2.15-m CASLEO 
telescope at San Juan, Argentina (see \cite{m99} for details), and low resolution CCD 
spectra obtained from the Vainu Bappu Observatory, India (see e.g. \cite{am99} for 
details).

\section{Results and discussion}

\begin{figure} \includegraphics[width=\textwidth]{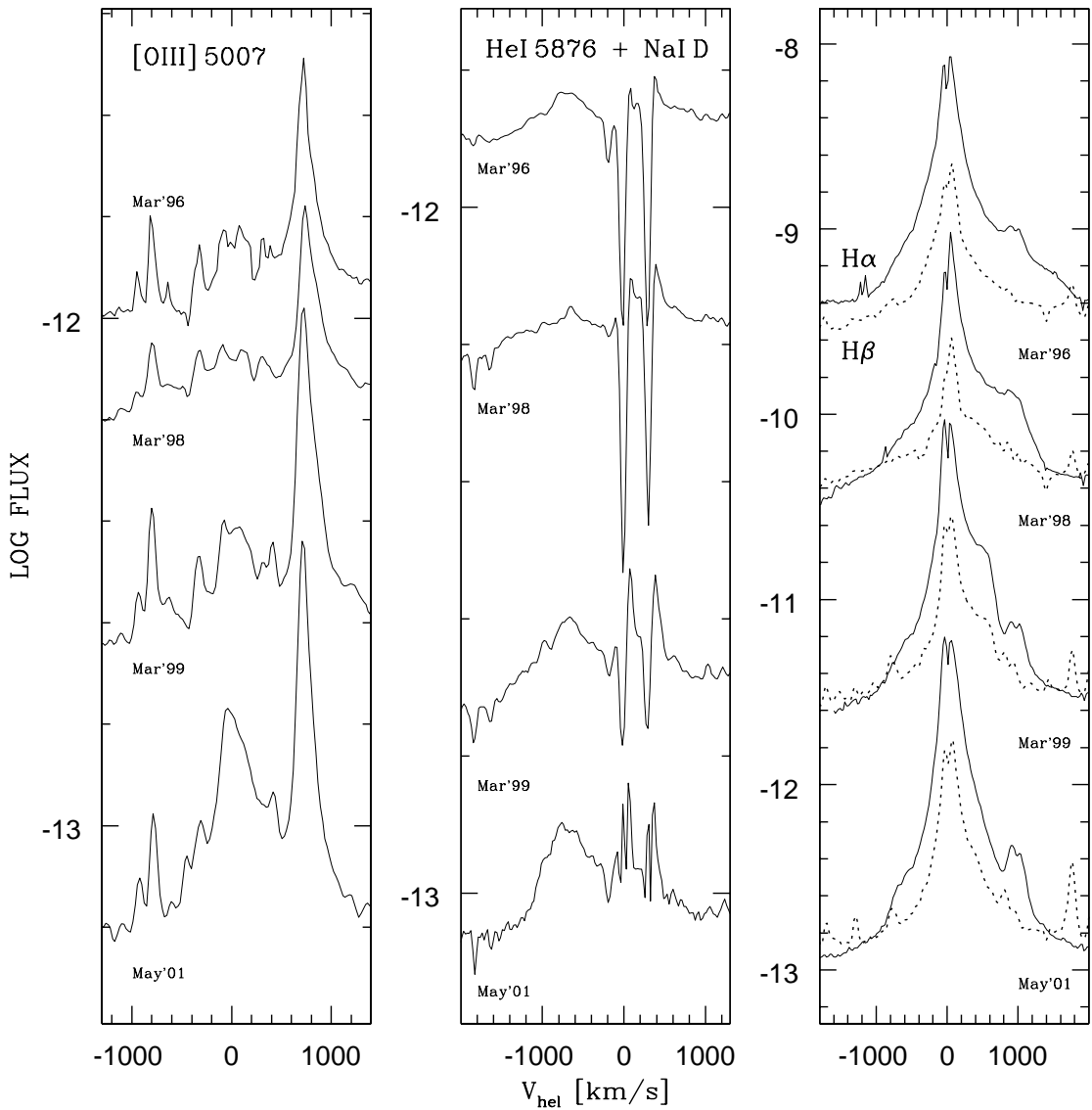} \caption{Evolution of 
emission line profiles in RX Pup in 1996$-$2001. The profiles are shifted vertically for 
better display.} \end{figure}

In the period covered by our spectroscopic observations, both components of RX Pup were 
active, although their changes were generally not correlated (Fig. 1). In particular, the 
near-IR flux was gradually decreasing in 1995$-$2000, suggesting that the Mira had entered 
a new obscuration phase (\cite{m01}).  At the same time, the visual light curve revealed a 
small eruption with maximum in 1998. The lack of any molecular absorption bands in our 
spectra ($\lambda \leq 8600\, \rm \AA$), as well as the absence of any Mira 
pulsations in the optical and red light curves (Fig.1; \cite{m99}; \cite{m01}) implies 
that the hot component is responsible for all changes in the optical and red spectral 
range. 

The optical brightening was accompanied by significant spectroscopic changes. The changes 
were the most remarkable in Balmer H\,{\sc i} lines: they were strong and narrow in $1988- 
91$ (\cite{m99}) whereas in $1995-2001$ they  developed a prominent flattened red wing 
(Fig. 2). The broad asymmetric wings were also present in He\,{\sc i} and near infrared 
Ca\,{\sc ii} triplet. The ionization level remained low, although the decline from the 
visual maximum in 1998 was associated with increase in [O\,{\sc iii}] and He\,{\sc i} 
emission line fluxes. Our spectra also showed the presence of a variable blue continuum 
with A/F-type shell absorption lines. The colour temperature of the continuum is low, 
$\sim 6000-9000\, \rm K$ for $E_{\rm B-V} \sim 0.8$, and it has a luminosity of $\sim 500- 
1000\, (d/1.8\,\rm kpc)^2\,\rm L_{\odot}$. On the other hand, the optical emission lines 
indicate a higher, $\sim 2.5-4 \times 10^4\, \rm K$, temperature source with a roughly 
comparable luminosity. \cite{m01} also detected large changes in the degree of 
polarization in the optical and red spectral range, and found that the polarized component 
is radiation from the blue continuum source scattered in the dust envelope surrounding the 
Mira.

We find here a striking similarity between RX~Pup and the quiescent symbiotic recurrent 
novae RS~Oph and T CrB, as well as the hot component of CH Cyg during bright phases. In 
all these systems, the optical data indicate the presence of relatively cool F/A- (CH Cyg) 
or A/B-type source (RS~Oph and T CrB), while the optical emission lines indicate a higher 
temperature source, with a roughly comparable luminosity (\cite{m88}; \cite{d96}; 
\cite{am99}). For example, in RS~Oph the IR colours indicate the presence of an additional 
warm, $\geq 7000$ K, source whereas the IUE and optical spectra show an A-B type shell 
source with $L \sim 100 - 600 \rm L_{\odot}$, accompanied by strong H\,{\sc i} and 
moderate He\,{\sc i} emission lines (\cite{d96}, and references therein). Similar variable 
UV/optical continuum with $L \sim 40-100 \rm L_{\odot}$ was observed in T~CrB (\cite{am99} 
and references therein). Although both in the symbiotic recurrent novae, T~CrB and RS~Oph, 
and in RX~Pup, the average luminosity of the B/A/F-type shell source is consistent with 
the accretion rate, $\dot{M} \geq 10^{-8}\, \rm M_{\odot}\,yr^{-1}$, required by the 
theoretical models, the effective temperatures places the hot components far from the 
standard massive white dwarf tracks in the HR diagram. Simultaneously, the X-ray data 
suggest 1$-$2 orders of magnitude lower accretion rates (\cite{m99}, and references 
therein). All three systems show similar quiescent behaviour: their hot components have 
highly variable luminosity and occasionally display blue-shifted absorption features, and 
broad asymmetric emission line wings (see also \cite{m99}; \cite{am99}; \cite{d96}).

We note here that similar A/F-type shell absorption spectrum was detected in AR 
Pav  and other Z And-type symbiotic stars with multiple outburst activity(e.g. 
\cite{q02}).  The blue absorption system in these systems traces the 
orbit of the hot component, and it is probably formed in a geometrically and 
optically thick accretion disc seen nearly edge-on and in a  gas stream 
(\cite{mk92}; \cite{q02}, and references therein). 
Parenthetically, the timescales and amplitudes for their eruptions 
are very similar to the timescales and amplitudes of the hot component 
luminosity variations (high and low states) in RX Pup and other symbiotic 
recurrent novae between their nova outbursts, as well as in accretion-powered 
systems CH Cyg and MWC 560.

It is also interesting that the shell spectrum appears only during the late decline from 
the nova outburst. In particular, the optical/visual light curves from the outbursts of 
T~CrB, RS~Oph and RX~Pup show more or less pronounced minima followed by a standstill (or 
secondary maximum) associated with the appearance of the shell spectrum. We suggest the 
minima are due to a decline in the hot component's luminosity after it passes the turnover 
in the HR diagram. The strong hot component wind during the plateau phase prevents 
accretion onto the hot component. Following the decline in luminosity the wind also 
ceases, and accretion of the material from the cool giant can be restored. We believe that 
the shell-type features and variable "false atmosphere" observed at quiescence together 
with the complex and broad emission lines originate from the accretion flow. The observed 
variability could be due to fluctuations in the mass-loss rate from the cool giant, the 
shell spectrum becomes stronger as $\dot{M}_{\rm cool}$ increases and as a result 
$\dot{M}_{\rm acc}$ increases. In the recurrent novae T CrB, RS Oph and the accretion-
powered systems of CH Cyg and MWC 560, the brightening of the hot component (or high 
activity stage) is associated with the presence of flickering (e.g. \cite{am99}). 
Such flickering, however, has not yet been detected in RX Pup.

Summarizing, the spectral development of RX Pup during the large outburst in 1969$-$1990 
resembles that of symbiotic novae AG Peg, HM Sge and V1329 Cyg as well as of the symbiotic 
recurrent nova RS Oph (\cite{m99}), although these other systems seem to be evolving on 
very different timescales, possibly reflecting differences in the mass of the white dwarf 
component. At quiescence the hot component shows activity (high and low activity states) 
resembling  that of symbiotic recurrent novae T CrB and RS Oph, which is probably due to 
relatively high mass transfer/accretion rate, of order of $\sim 10^{-7}\, \rm 
M_{\odot}\,yr^{-1}$.  A high accretion rate, $\geq 10^{-8}\, \rm M_{\odot}\,yr^{-1}$, is 
also required to account for the recurrence time, $\sim$ 80, 80 and 22 yr, for RX Pup,  
T CrB, and RS Oph, respectively.

%%%%%%%%%%%%%%%%%%%%%%%%%%%%%%%%%%%%%%%%%%%%%%%%
%% BACKMATTER
%%%%%%%%%%%%%%%%%%%%%%%%%%%%%%%%%%%%%%%%%%%%%%%%

\begin{theacknowledgments}
  This study was partly supported by KBN research grant No. 5P03D 019 20.
\end{theacknowledgments}

%% If the bibliography is produced without BibTeX comment out the

\bibliographystyle{aipprocl} % if natbib is missing

\begin{thebibliography}{}

\bibitem{m99} Miko{\l}ajewska, J., Brandi, E., Hack, W., Whitelock, P.A., Barba, R., 
Garcia, L., Marang, F., 1999, MNRAS, 305, 190 

\bibitem{am99} Anupama, G.C., Miko{\l}ajewska, J., 1999, A\&A, 344, 177 

\bibitem{m01} Miko{\l}ajewska, J., Brandi, E., Garcia, L., Ferrer, O.,  W., Whitelock, 
P.A.,  Marang, F., 2001, in Szczerba R. et al., eds, Post-AGB Objects as a Phase of 
Stellar Evolution, Kluwer, 227  

\bibitem{m88} Miko{\l}ajewska, J., Selvelli, P.L., Hack, M., 1988, A\&A, 198, 150

\bibitem{d96} Dobrzycka, D., Kenyon, S.J., Proga, D., Miko{\l}ajewska, J., Wade, R., 1996, 
AJ, 111, 2090

\bibitem{q02} Quiroga, C., Miko{\l}ajewska, J., Brandi, , Ferrer, O., Garcia, L., 2002, 
A\&A, 387, 139

\bibitem{mk92} Miko{\l}ajewska, J., Kenyon, S.J., 1992, AJ, 103, 579 \end{thebibliography}


\IfFileExists{\jobname.bbl}{}
 {\typeout{}
  \typeout{******************************************}
  \typeout{** Please run "bibtex \jobname" to optain}
  \typeout{** the bibliography and then re-run LaTeX}
  \typeout{** twice to fix the references!}
  \typeout{******************************************}
  \typeout{}
 }

\end{document}